\documentclass[showkeys]{revtex4}

\usepackage{epsfig}

\begin{document}

\title{Numerical Analysis of the Stub Transistor}
\author{ Alexandre B. \ Guerra and Edval J.\ P.\ Santos}
\affiliation{Laboratory for Devices and
Nanostructures at the Departamento de Eletr\^onica e Sistemas,
Universidade Federal de Pernambuco,
Caixa Postal 7800, 50670-000, Recife-PE, Brazil\\
E-mail: edval@ee.ufpe.br .
}

\begin{abstract}
Stubbed waveguides and stub transistors are
candidates for next generation electronic devices.  In
particular, such structures may be used in spintronics-based
quantum computation, because of its ability to induce spin-polarized 
carriers. In this paper, we present the simulation of the
conductance of the stub transistor (single and double-gated), 
modeled with a nearest-neighbor  tight-binding Hamiltonian.  
The oscillatory behavior of the channel conductance with the 
applied stub voltage is observed.

\end{abstract}

\keywords{mesoscopic, stub transistor, simulation.}

\maketitle

\section{Introduction}
Many devices are being proposed to substitute the MOSFET,
as microelectronics reaches the nanoscale. Among them are
the stubbed waveguide and the stub 
transistor~\cite{Sols1988,Sols1989,DJ1994,DM2000,DRVRPM2000,WVP2002}.  
The stubbed waveguide  consists of a nanowire with a set of 
periodically spaced stubs. The stub transistor is a nanowire 
with an electric potential applied to the stub, which acts 
as a gate voltage, see Fig.~1.  Such devices can be fabricated with 
dimensions below $100 nm$, which,  at low temperatures, is 
smaller than the length scale over which the electron preserves 
its phase coherency. This length is known as the electron 
phase-coherence length, $L_{\phi}$, and the transport 
through the device is ballistic. These kind of structures
are classified  as mesoscopic systems and cannot be described 
by the usual semi-classical transport theory,  the wave nature of 
the electron needs to be taken explicitly into account.

The conductance, $G$, of the stub transistor
displays an oscillatory behavior, as the electric
potential applied to the stub, $V_G$ is varied.  
Therefore, the channel can be
opened and closed.  A minimum of $G$ is a reflection
resonance or antiresonance and a maximum of $G$ is a
transmission resonance or just resonance~\cite{DJ1994}.  
The transmission zeroes occur as function
of the length of the stub, $L_{stub}$, given by $kL_{stub}= n\pi$,
and the resonances are given by  $kL_{stub}= (n + 1/2)\pi$.
The use of multiple periodically placed stubs transforms
the transmission zeroes in blocked transmission bands~\cite{DJ1994}.
Such devices have been fabricated on GaAs/AlGaAs by
MBE~\cite{DRVRPM2000}.

To understand the behavior of the device, many theoretical
techniques have been used, such as, the recursive Green function 
technique~\cite{Sols1989} to calculate the scattering matrix, 
the transfer matrix technique~\cite{DJ1994,WVP2002}. 
Santos~\cite{Edval2002} has developed a recursive Green 
function program to simulate ballistic quantum devices.

The reason for the great interest in the stub transistor is
the potential for ultrafast signal processing without giving
prohibitive energy dissipation. The traditional FET transistor
works using a ``brute'' force switching voltage, because the electric
field has to deplete electrons under its gate to change the
conductance. The quantum interference devices, QIDs, on the other 
hand, work by changing the electron interference pattern.  This 
quantum phenomenon requires a lower switching voltage, therefore 
a lower energy dissipation.  The reduced size of the QIDs decreases 
the time that the electron takes to travel through it, as a consequence, 
its switching frequency can be increased up to the terahertz 
frequency region~\cite{MKS2000}.

Strong analogies exist between ballistic quantum devices and 
similar devices in optics and microwave engineering. One, however, 
has to take care when try to implement optical or microwave devices 
using matter electronic waves. These analogies are not perfect 
because electrons  interact strongly and obey Fermi statistics,
while photons are bosons.

In this paper, the quantum stub transistor is examined by applying
the recursive Green function technique.  The paper is divided
in five sections, this introduction is the first, next the
quantum stub transistor is described. Third an overview of the
simulation method is presented. Next, the results and analysis 
are presented. Finally, the conclusions.

\section{The quantum stub transistor}

\begin{figure}
        \epsfxsize=6in
\vskip 2in
        \centerline{\epsffile{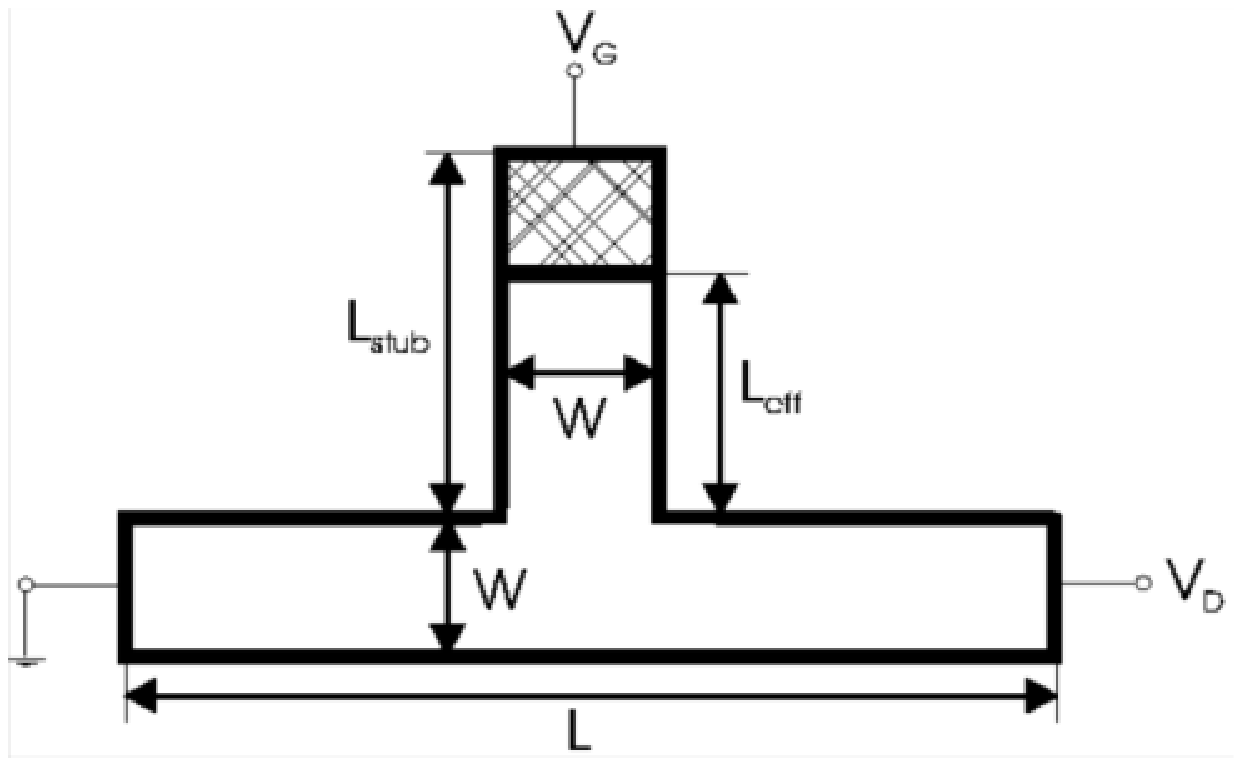}}
        \label{esquema}
        {\Large Fig .1 - Guerra and Santos}
\end{figure}

A schematic view of the quantum stub transistor is shown in
Fig.~\ref{esquema}, where $V_{G}$ is the gate voltage and $V_{D}$
is the drain voltage. The drain voltage is related to the incident 
electron energy by $E= eV_{D}$. If the stub length, $L_{stub}$, 
is much larger than the inelastic mean free path $L_{ine}$, then this 
structure represents just two intersecting wires and the transconductance, 
$g_{m}= {\partial I_{D}/\partial V_{G}}$ is zero, where $I_{D}$ is the 
drain current. For $L_{stub}\leq L_{ine}$, however, the quantum 
nature of the electronic transport
makes the transconductance $g_{m}$ different from zero and the
situation changes completely. What happens is similar to the
conventional metal waveguide, where one has to move a piston
inside a cavity to change the interference patterns of the
electromagnetic field. In the case of the quantum stub transistor,
the gate voltage  acts as a piston by varying
the length of the stub. This changes the penetration of the electron 
inside the stub, and therefore the interference pattern. To represent
this change of the stub length with the gate voltage, $L_{eff}$
is defined. The stub
length $L_{stub}$ for which quantum effects become important vary
widely depending on temperature and material used. In ultra-pure
semiconductor material and at liquid Helium ($4.2 K$) temperatures
and below, coherent effects are observable over distances of 
$1 \mu m$~\cite{Tpalm1993}.

The confinement of the propagating  electrons from source to drain
in the transverse direction allows only discrete energy levels or
modes. The number of transverse modes, $M$, for a quantum stub transistor
with width $W$ is given by the ratio between the width and the
Fermi wavelength, $\lambda_{F}$, as shown in Equation~1.

\begin{equation}\label{relacao}
M=Int\left(\frac{W}{\frac{\lambda_{F}}{2}}\right)
\end{equation}
where $Int(x)$ represents the integer that is just smaller than $x$. 

If more than one mode
propagates through the device, each mode will have a different penetration 
length in the stub, as a consequence the device will display an erratic 
interference pattern. Hence, for a well behaved device the  incident energy, 
$E$, must be in the range $E_{1}\leq E< E_{2}$, where $E_{1}$ and $E_{2}$ 
are the energy of the first and second mode respectively. The energy of 
the $n$-th mode can be estimated with the infinite potential box 
approximation.

\begin{equation}\label{energia}
E_{n}=\frac{\hbar^{2}}{2m^{*}}\frac{n^{2}\pi^{2}}{W^{2}}
\end{equation}
where $m^{*}$ is the electron effective mass. 

The electron may follow many different paths, as it propagates
through the device.  The maximum length difference occurs between
a path going straight from source to drain and  another going
through the stub and then to the drain. This difference is 
approximately $2L_{stub}$. So the
number of minima expected in the interference pattern  is $m+1$,
where $m$ is the maximum integer that satisfies the inequality below
\begin{equation}\label{desigualdade}
L_{stub}\geq(m+\frac{1}{4})\lambda_{F}.
\end{equation}
For example, for  $L_{stub}= 30nm$ and $\lambda_{F}= 20nm$, one
gets $m+1= 2$.

\section{The simulation method}

To simplify the analysis, it is assumed that the potential inside the 
device is equal to zero, and the borders of the device are defined
by a high potential barrier. Even with these simplified boundary 
conditions, the two dimensional nature of the devices makes the problem 
of finding the total transmission probability, $T$, of the electron through 
the device, analytically intractable.  To obtain the conductance of the 
device using the Laudauer formula ($G=\frac{2e^{2}}{h}T$), it is necessary
to use a numerical method to solve  Schr\"{o}dinger equation 
($H\Psi=E\psi$). Hence, the geometry has to be described with a model
Hamiltonian, this is done, by using the nearest-neighbor tight-binding  
model. To implement such model, the structure showed in Fig.~\ref{esquema} 
is filled with a sufficient dense square lattice with periodicity $a$.
The plane wave motion can be emulated by a tight-binding Hamiltonian, 
when the lattice parameter, $a$, is much smaller than all other length 
scales of the problem. Proceeding  as indicated, one may reasonably 
expect to correctly reproduce the continuous dynamics.

\begin{figure}
        \epsfxsize=7in
\vskip 2.5in
        \centerline{\epsffile{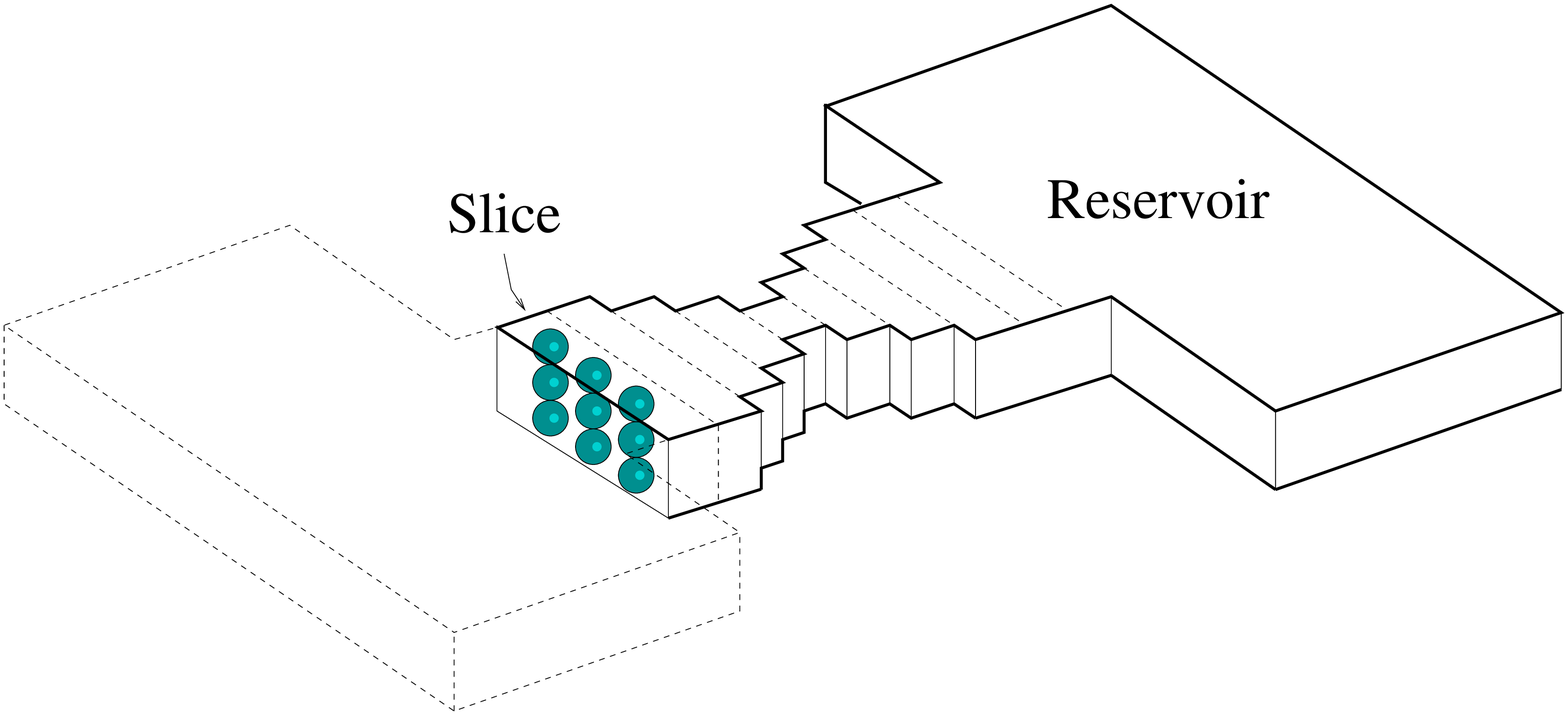}}
	\label{nanoestrutura}
        {\Large Fig .2 - Guerra and Santos}
\end{figure}

  To better illustrate the method, an example of a square lattice is
shown in Fig.~\ref{nanoestrutura}.  Each slice is stored as a
column of the device matrix, where the matrix entry is related to
the potential energy confining the electron. Each column (slice)
of the device matrix is used to construct the Hamiltonian or slice 
matrix for numerical calculation by using the tight-binding Hamiltonian
Model. The $n$th slice matrix is shown in Equation~4.

\begin{equation}
H_n= \left(
\begin{array}{ccccc}
V_1 - 4t & -t   & 0      & ... & 0     \\
-t     & V_2 - 4t & -t & ...   & 0     \\
0      & -t   & V_3 - 4t & ... & 0     \\
.      &   .    & .      &     & .     \\
.      &   .    & .      &     & .     \\
.      &   .    & .      &     & .     \\
0      &   0    & 0      & ... & V_M - 4t
\end{array}
\right)
\end{equation}where $V_i$ in the diagonal is the local potential
energy profile at a given slice, and $t$ is a coupling term, given by
Equation~5.

\begin{equation}\label{parameto}
t=E_{F}\left(\frac{1}{2\pi}\right)^{2}\left(\frac{\lambda_{F}}{a}\right)^2
\end{equation}
where $E_{F}$ is the Fermi energy level. 

The  Schr\"{o}dinger equation is now solved by using the Green function 
method, i.e., the equation to the solved has the form:
\begin{equation}\label{Green}
(E\underline{I}-\underline{H})\underline{G}= \underline{I}
\end{equation}
where $\underline{I}$ is the identity matrix, and $\underline{G}$ 
is the Green function matrix. For simulations with practical interest, 
the direct inversion of the matrix $(E\underline{I}-\underline{H})$ requires 
huge amounts of memory and  processing power. This is circumvented by
employing an iterative method for the  computation of the Green function. 
The details of the program  can be found in Reference~\cite{Edval2002}.

The data input process is time consuming, as the device has to be 
described in matrix format. To facilitate data input, a front-end 
that converts the device geometry from a standard bit-mapped format 
into the matrix format that the computation algorithm understands has 
been developed~\cite{GS2002}. The relation between the 
width $W$ of the device and the number of pixels $P_{W}$ used to 
represent this dimension is given by

\begin{equation}\label{comprimento}
W=(P_{W}+1)a.
\end{equation}
For $a= 0.5nm$ and $W= 10 nm$, $P_W= 19$.

\section{Results and analysis}

The program uses normalized units, $E_{F}$ is the unit of energy and
$2e^{2}/{h}$ is the unit of conductance.
In the simulations, it is assumed that the gate voltage, $V_{G}$,
drops linearly along the stub\footnote{The program does not have a Poisson
solver for the calculation of the gate potential distribution, this
feature will be added in the near future.}. 
The effective mass is $m^{*}=0.07m_{0}$ (GaAs),
the wire width is $W= 10 nm$ and the lattice parameter is
$a= 0.5 nm$.

\begin{figure}
        \epsfxsize=7in
\vskip 2.5in
        \centerline{\epsffile{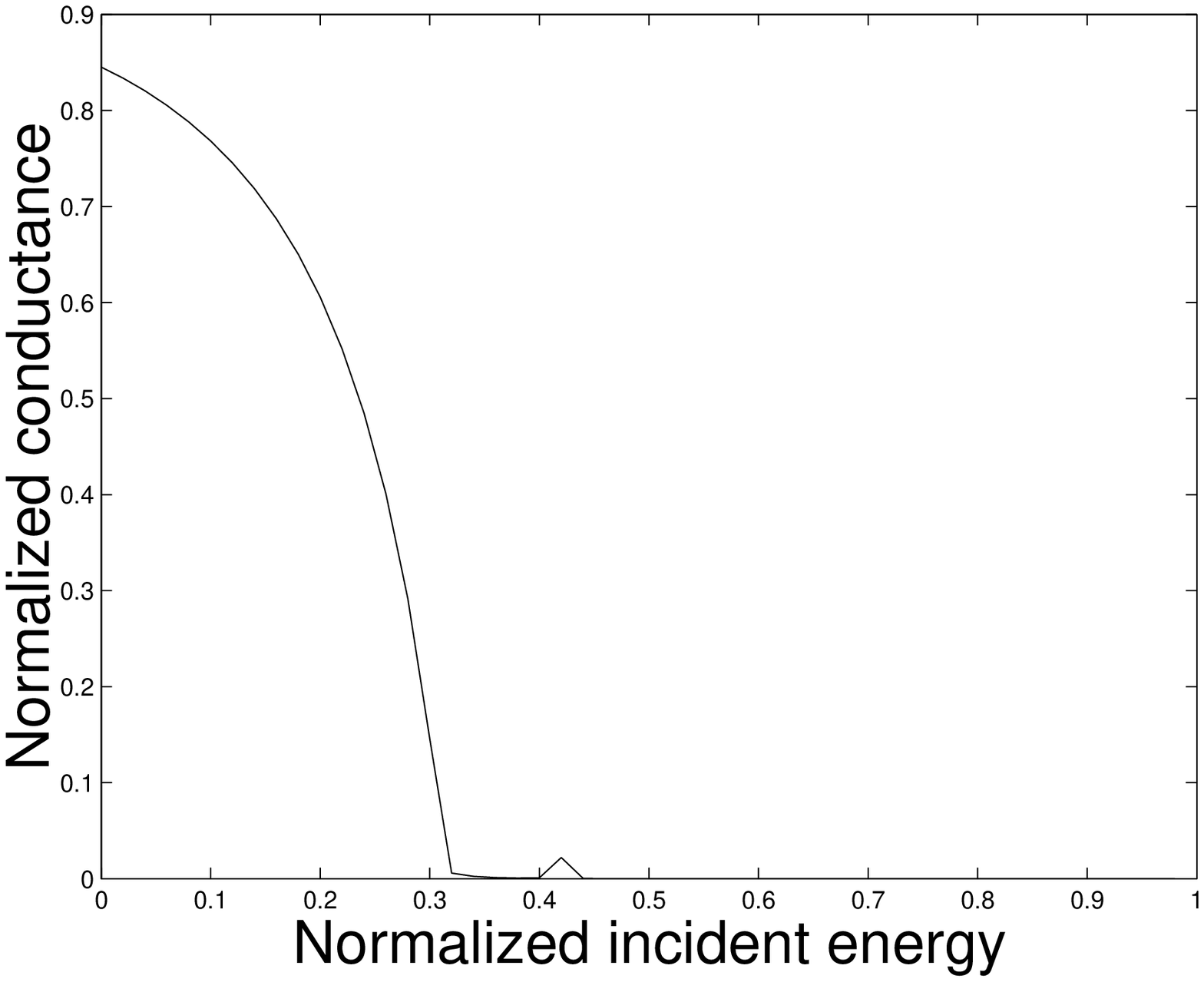}}
        {\Large Fig .3 - Guerra and Santos}
	\label{modo}
\end{figure}

The normalized coupling term is calculated with  Equation~\ref{parameto}.
Considering the single mode device, $t= 40.52$. Using an 
incident energy equal to the first allowed mode $E_{1}= 53meV$,
the plot of the conductance as a function of the normalized incident
energy is presented in Figure~\ref{modo}. The result confirms 
that the device has only one mode. The gate voltage is set 
equal to zero, $L= 73.3 nm$ and $L_{stub}= 11 nm$.

\begin{figure}
        \epsfxsize=7in
\vskip 2.5in
        \centerline{\epsffile{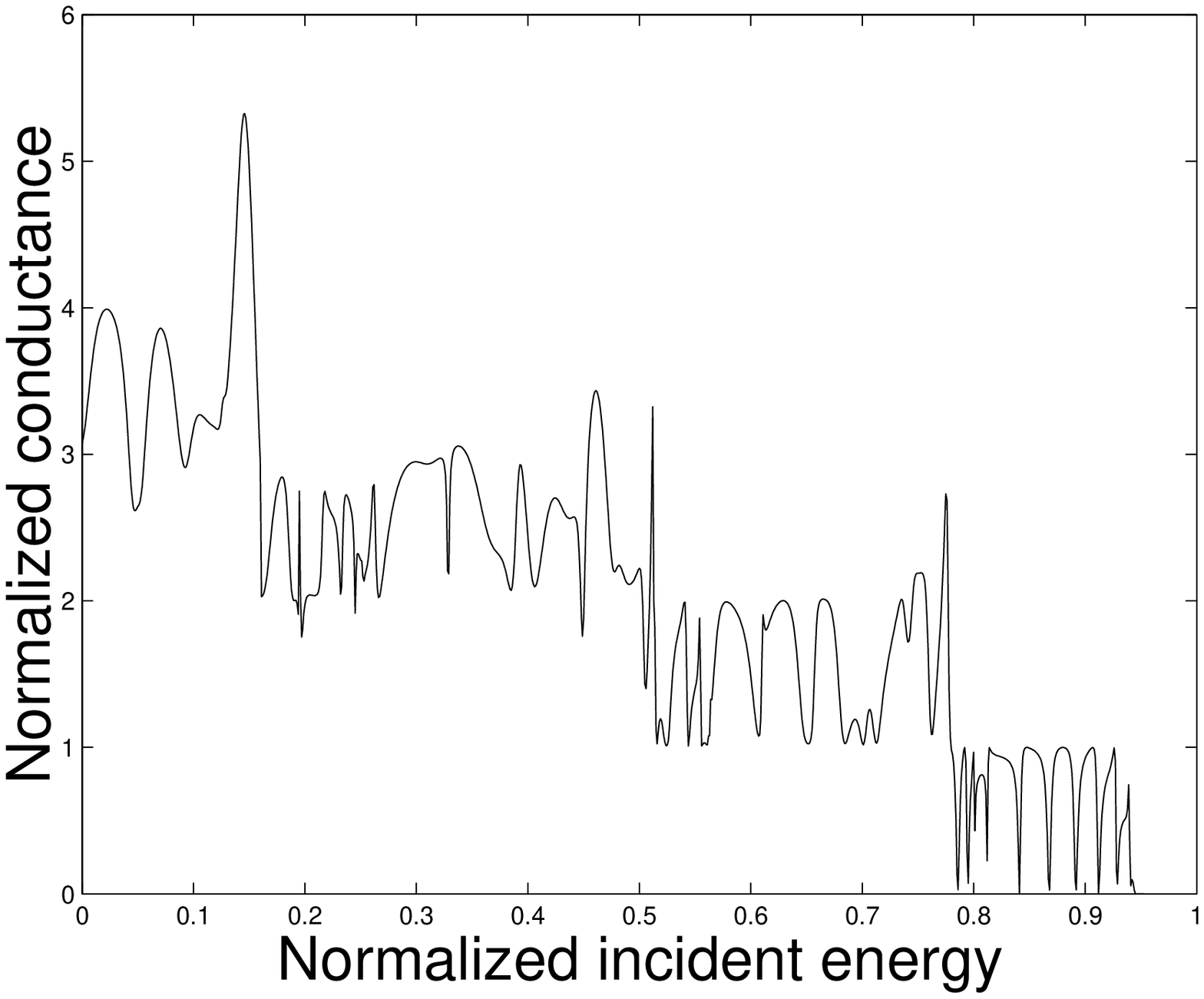}}
        {\Large Fig .4 - Guerra and Santos}
	\label{fourmodes}
\end{figure}

Considering now a device with four open transmission channels,
the simulation yields a conductance with interference patterns,
at each conductance step~\cite{DRVRPM2000}. The plot of the normalized 
conductance as a function of the normalized incident energy
is presented in Figure~\ref{fourmodes}. For this simulation,
$W= 10 nm$, $P_W= 19 pixels$, $L_{stub}= 50 nm$, $M= 4$ (four modes),
and $t= 2.533$.

\begin{figure}
        \epsfxsize=7in
\vskip 2.5in
        \centerline{\epsffile{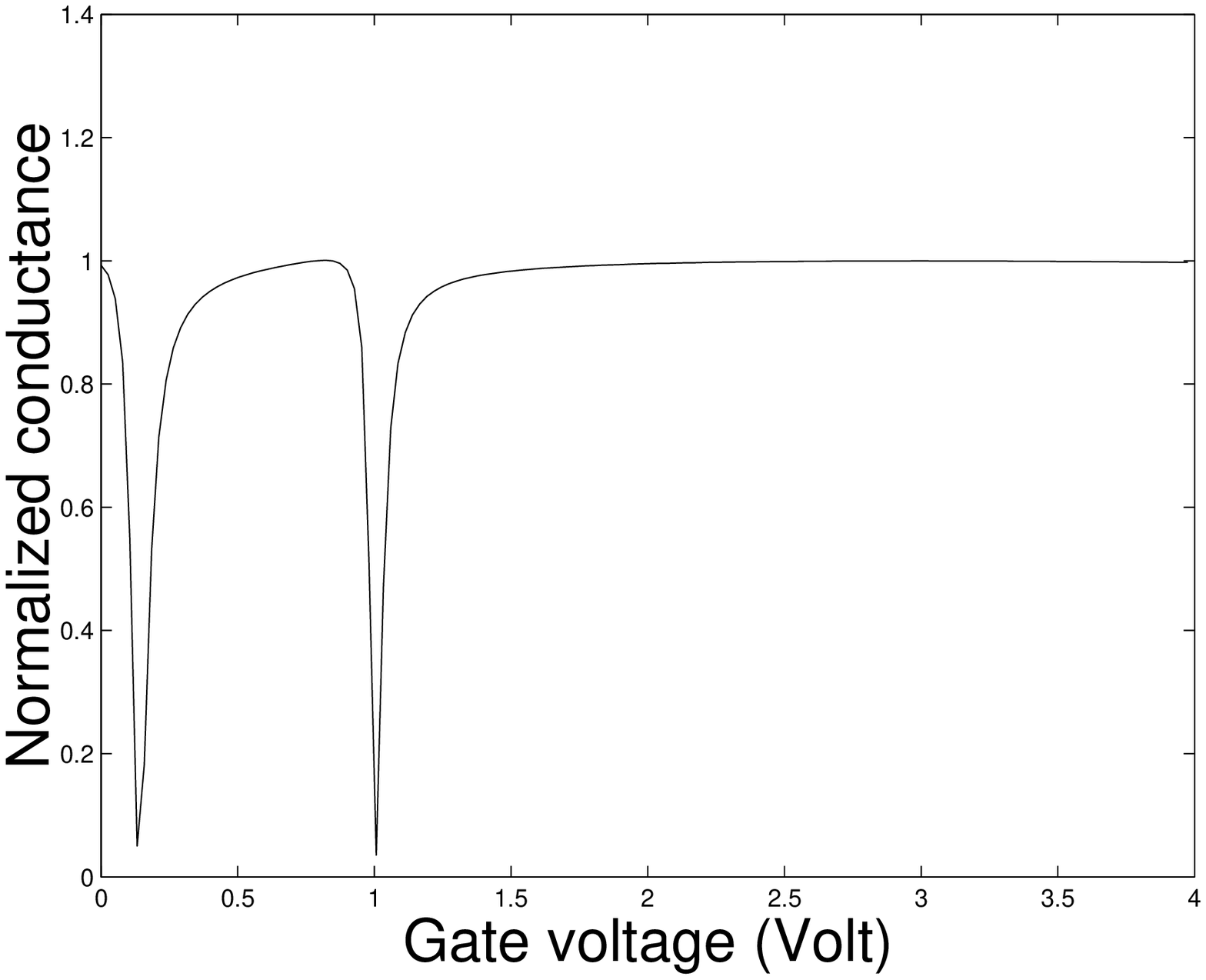}}
        {\Large Fig .5 - Guerra and Santos}
	\label{min1}
\end{figure}

\begin{figure}
        \epsfxsize=7in
\vskip 2.5in
        \centerline{\epsffile{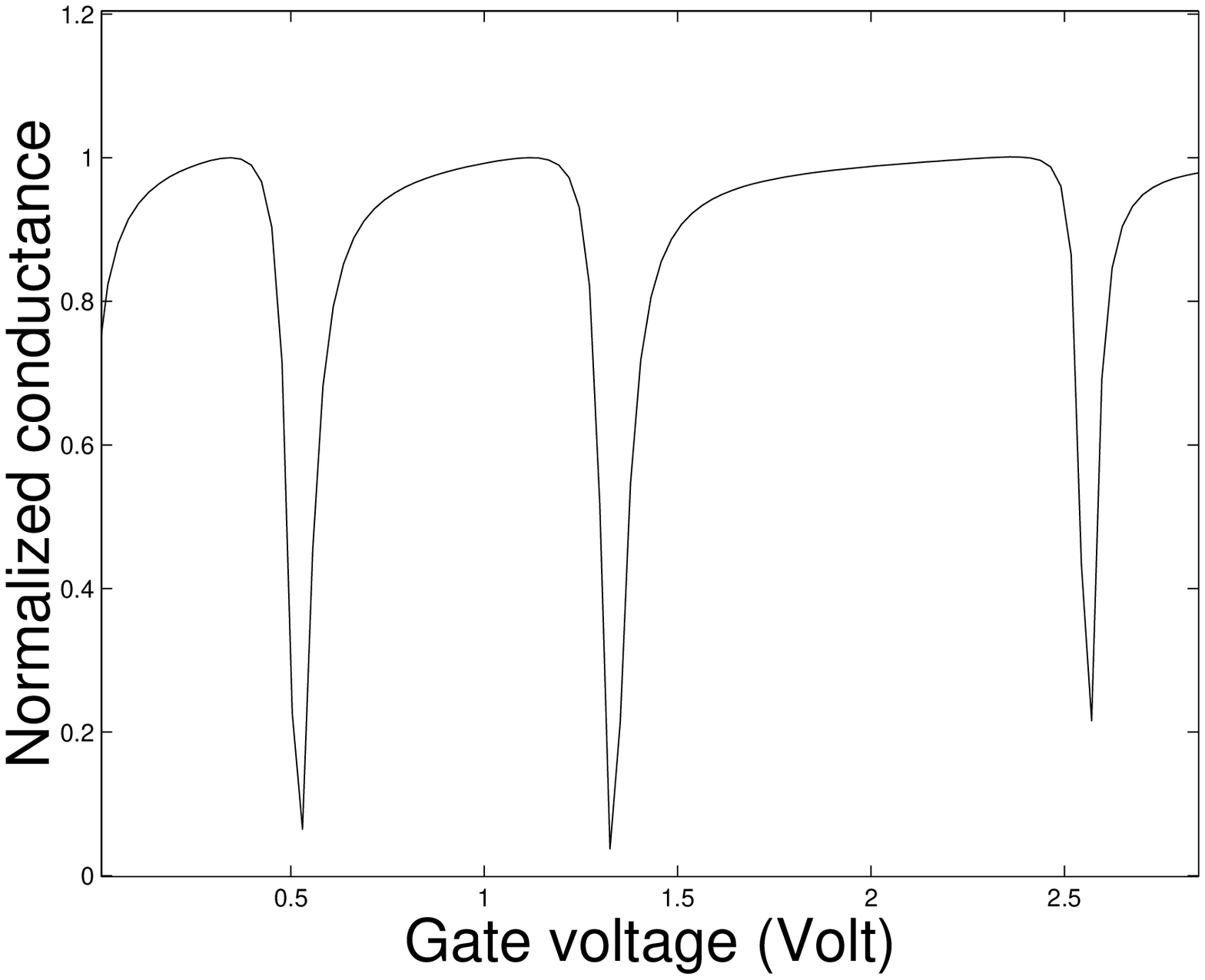}}
        {\Large Fig .6 - Guerra and Santos}
	\label{min2}
\end{figure}

Next, the incident energy (drain voltage) is fixed at 
a position to get just one open channel, and the gate voltage is 
varied. The normalized conductance as a function of the gate
voltage are presented in Figure~\ref{min1} and \ref{min2}.
The results show that the conductance varies between one
and zero, as the gate voltage varies. In Figure~\ref{min1},
there are two conductance dips or antiresonances 
(zero conductance). In Figure~\ref{min2}, there are
three antiresonances. This can be used to test
the correctness of Equation~\ref{desigualdade}, 
two values $L_{stub}$ are used. For this calculation,
the incident energy is $53 meV$ and $\lambda_F= 20 nm$. 
Hence, for $L_{stub}= 29 nm$, one gets $m= 1$, and the
number of minima is $m+1= 2$. For  $L_{stub}= 50 nm$, one 
gets $m= 2 $, and the number of minima is $m+1= 3$.
The calculated number of minima are in accordance with 
the simulation.

\begin{figure}
        \epsfxsize=6in
\vskip 2.5in
        \centerline{\epsffile{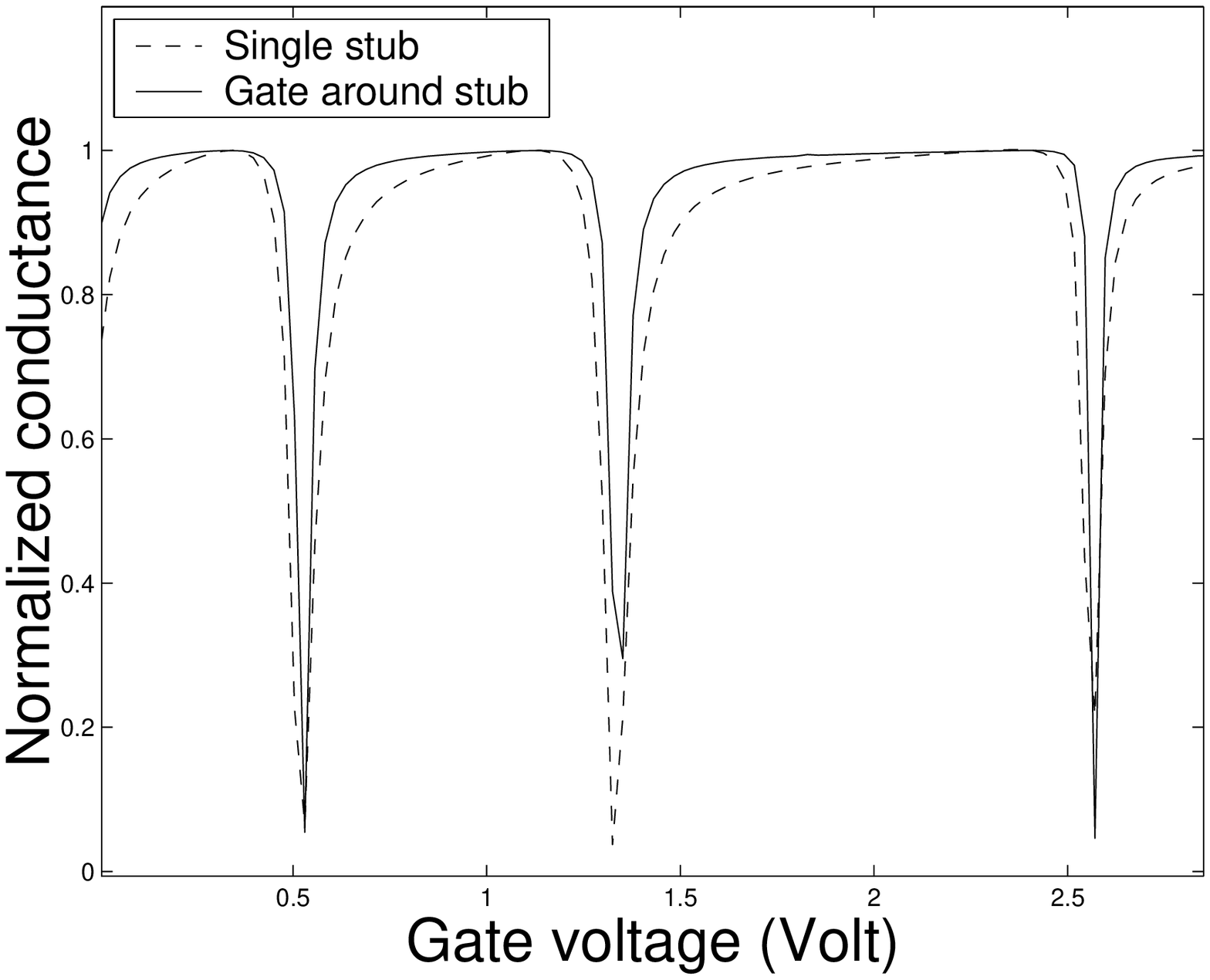}}
        {\Large Fig .7 - Guerra and Santos}
	\label{gatearound}
\end{figure}

Another configuration is the gate-all-around stub transistor.
This transistor has a double stub, one opposite to the other.
The calculated conductance as a function of the gate
voltage is presented in Figure~\ref{gatearound}. The presence 
of the gate-all-around stub narrows the antiresonance.

These are interesting results, because one can control the channel 
transmission, by applying a control voltage, which is useful in 
digital logic, for instance. The main practical difficulty is that 
the width of the antiresonance is very narrow, and one would have 
to set the voltage with great precision. However, It is known that 
in a stubbed waveguide the presence of a series of stubs transforms 
the antiresonance in blocked bands~\cite{DJ1994}.  The idea is to 
increase the number of stubs to get a wider blocked transmission band.
The simulation is now carried out varying the number of
stubs along the device.  The double stub transistor is shown in 
Figure~8. The result, presented in Figure~9, shows that the 
antiresonance is replaced by  double antiresonances. By adding
three stubs, one gets a triplet. Therefore, 
increasing the number of stubs may be a practical solution 
for the realization of stub transistors, as the blocked
transmittance gets less sensitive to a particular gate voltage level.

\section*{Conclusions}

Besides the potential for high-speed operation, the appeal of the
quantum stub transistor is its  very small dimensions and the
possibility of operation at very low power levels, which could
allow extremely high-level of integration. But, as pointed by
Laundauer, like the others (QIDs), this kind of device requires
very specific values of the control signal and of the devices
parameters for a conductance maximum or minimum. This leaves little
room for device fabrication errors. However, the results in this paper show
that the use of multiple stubs reduces this problem by widening the
antiresonance. This may be a reasonable approach for the practical 
realization of stub transistors. 

\section*{Acknowledgments}

The authors would like to acknowledge the support of CNPq
under the ``Instituto do Mil\^{e}nio'' Initiative.

\bibliographystyle{IEEE}

\newpage
\begin{center}
\large{ Figure captions }
\end{center}

Fig.~1. Schematic view of the quantum stub transistor, where
$V_G$ is the voltage applied to the stub, $V_D$ is the voltage
applied to the nanowire, $W$ is the width of the nanowire, 
$L_{stub}$ is the stub length, and  $L_{eff}$ is the 
stub effective length.

Fig.~2. Quantum device divided in slices for the simulation of
quantum transport.  The device is connected to two reservoirs.

Fig.~3. The quantum stub transistor with only one open transmission
channel. $W= 10nm$, $a= 0.5 nm$, $P_W= 19 pixels$,  $L_{stub}= 11 nm$, 
$M= 1$ (one mode), and $t= 40.52$.

Fig.~4. The quantum stub transistor with four modes.
$W= 10 nm$, $P_W= 19 pixels$, $L_{stub}= 50 nm$, $M= 4$ (four modes),
and $t= 2.533$.

Fig.~5. Normalized conductance as a function of the gate voltage $V_{G}$, 
for $L_{stub}=29 nm$.

Fig.~6. Normalized conductance as a function of the gate voltage $V_{G}$,
for $L_{stub}=50 nm$.

Fig.~7. Conductance of the gate-all-around stub transistor
as a function of the gate voltage $V_{G}$,
$L_{stub}=50 nm$ at each side.

Fig.~8. Geometry of the double stub stub transistor,
the stub separation is $10 nm$, $L_{stub}= 50 nm$. 

Fig.~9. Conductance of the double stub stub transistor
as a function of the gate voltage $V_{G}$,
the stub separation is $10 nm$, $L_{stub}= 50 nm$ at each side.

\newpage

\newpage

\newpage

\newpage

\newpage

\newpage

\newpage

\newpage

\begin{figure}
        \epsfxsize=6in
\vskip 2.5in
        \centerline{\epsffile{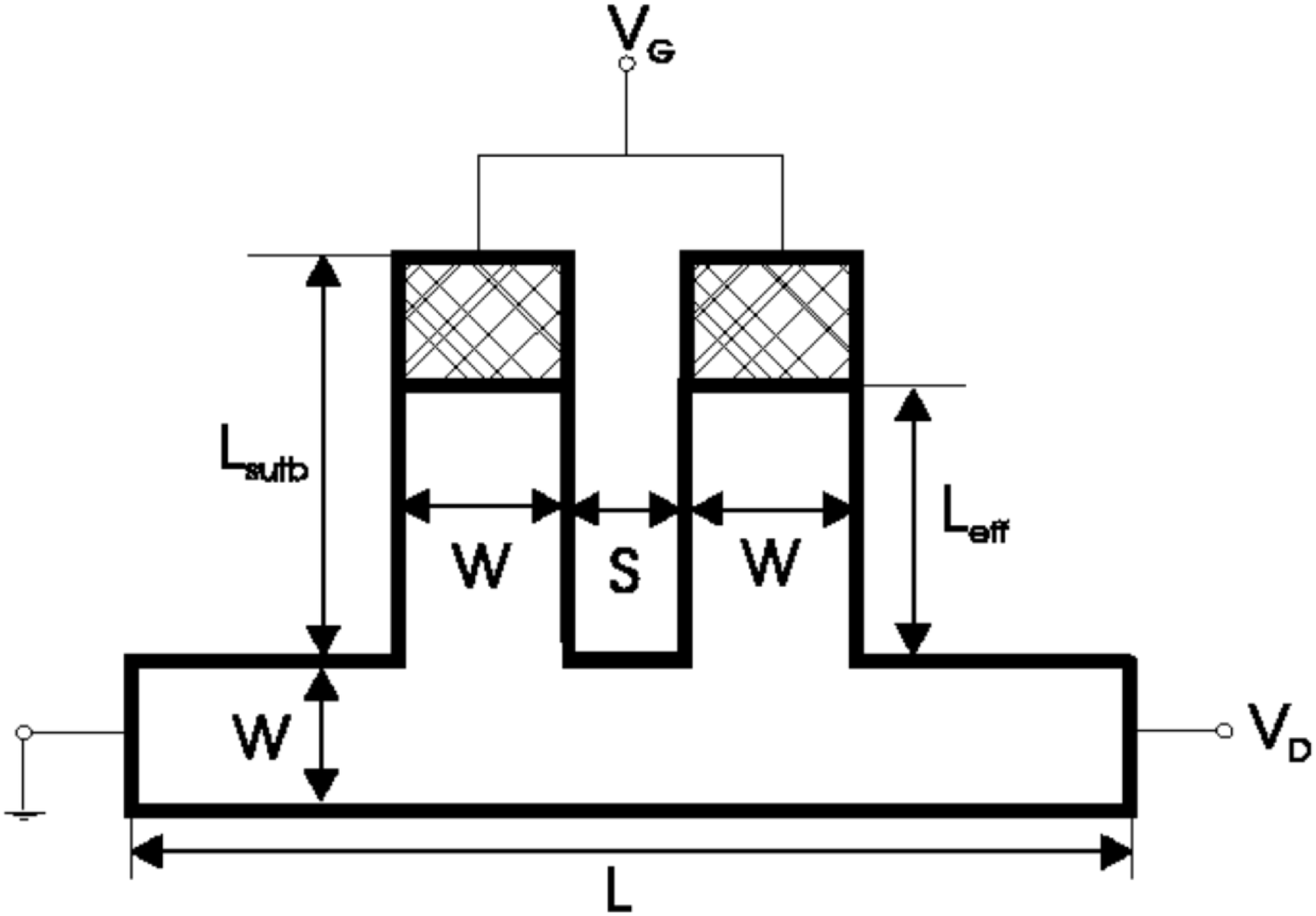}}
        {\Large Fig .8 - Guerra and Santos}
	\label{doublestub}
\end{figure}

\newpage

\begin{figure}
        \epsfxsize=6in
\vskip 2.5in
        \centerline{\epsffile{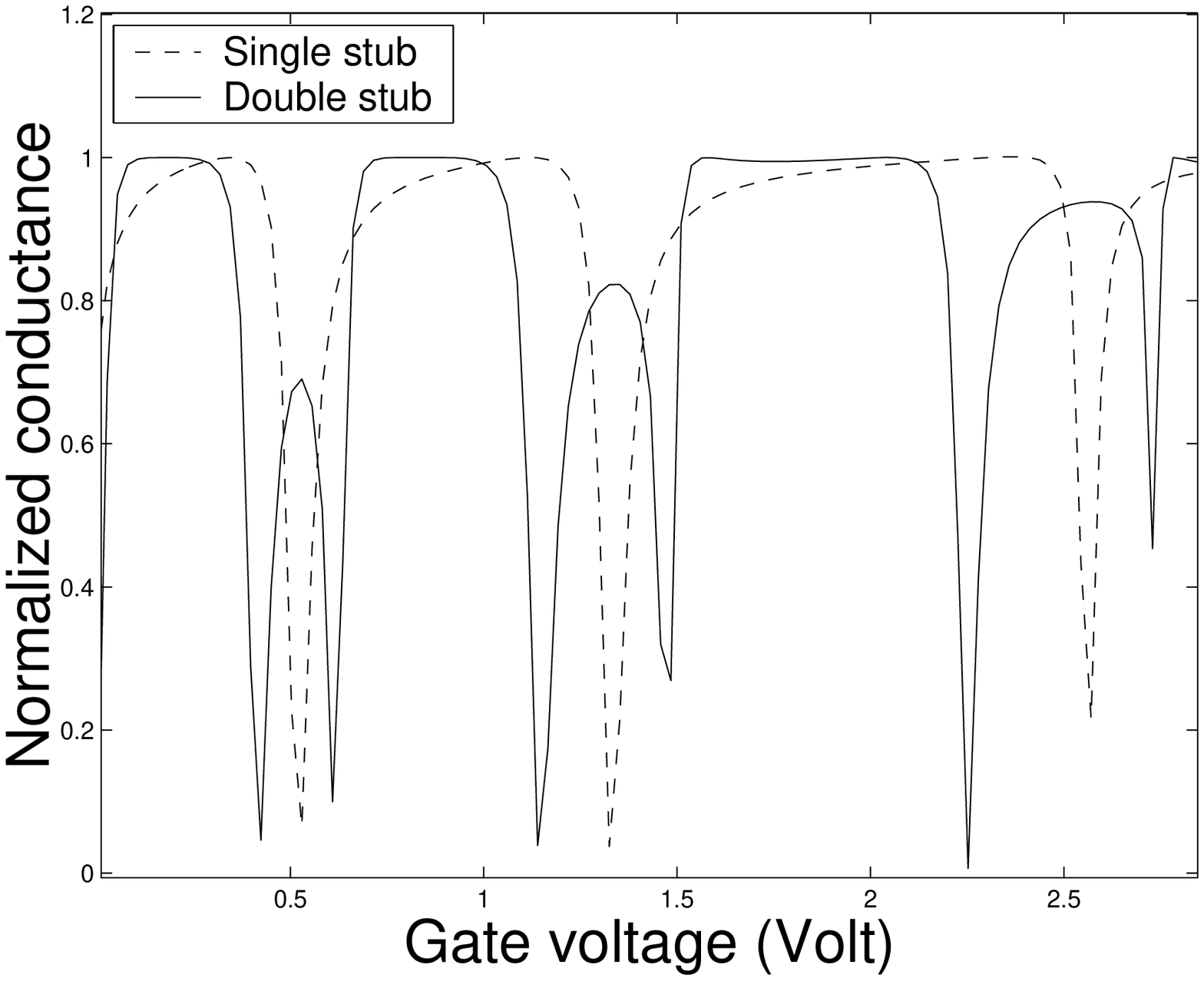}}
        {\Large Fig .9 - Guerra and Santos}
	\label{doublestub2}
\end{figure}

\end{document}